\documentclass[runningheads]{llncs}

\usepackage[final,year=2026]{eccv}

\usepackage{eccvabbrv}

\usepackage{graphicx}
\usepackage{booktabs}
\usepackage[toc]{appendix}

\usepackage[accsupp]{axessibility}

\usepackage{hyperref}

\usepackage{orcidlink}

\usepackage{amsmath,amsfonts,bm}

\def\eqref#1{equation~\ref{#1}}

\def\1{\bm{1}}

\DeclareMathAlphabet{\mathsfit}{\encodingdefault}{\sfdefault}{m}{sl}
\SetMathAlphabet{\mathsfit}{bold}{\encodingdefault}{\sfdefault}{bx}{n}

\newcommand{\E}{\mathbb{E}}

\newcommand{\R}{\mathbb{R}}

\newcommand{\vect}[1]{\bm{#1}}

\begin{document}

\title{Climate Physics Dynamic Matching\thanks{ECCV 2026 Workshop on AI for Climate and Conservation (AICC).}}

\author{Gurjeet Sangra Singh \inst{1,2}\orcidlink{0009-0008-2340-5867} \and
Frantzeska Lavda\inst{2}\orcidlink{0000-0003-2868-8380} \and
Alexandros Kalousis \inst{2}\orcidlink{0000-0001-6282-0686}}

\authorrunning{Singh, S. G. et al.}

\institute{University of Geneva, Geneva, Switzerland \and HES-SO Geneva, Geneva, Switzerland
\\
\email{ gurjeet.singh@etu.unige.ch \{frantzeska.lavda,alexandros.kalousis\}@hesge.ch}}

\maketitle

\begin{abstract}
Deep generative models such as flow matching and diffusion models have shown potential for learning complex dynamical systems, but typically act as black boxes that neglect underlying physical structure, while physics-based models governed by partial differential equations are often incomplete due to missing source terms, or uncertain parametrisations. We present Climate Physics Dynamic Matching (ClimPhyDM), a variational simulation-free dynamics informed framework for weather forecasting that combines an advection-type physics prior with data-driven components in a variational framework.
On the ERA5 benchmark at hourly (42-hour) and monthly (5-month) resolutions, ClimPhyDM outperforms ClimODE, and GB-DM, keeping the lower error at extended horizon, indicating improved temporal stability and resistance to error accumulation,  while its simulation-free paradigm also enables training on a single modest 12 GB consumer GPU.

\end{abstract}

\section{Introduction}\label{sec:intro}
Modelling the dynamics of the Earth's atmosphere is a central challenge in climate science and weather prediction. Deep learning approaches to weather forecasting have made remarkable progress in recent years. Models such as Pangu-Weather~\cite{bi2023accurate}, GraphCast~\cite{lam2023learning}, and FourCastNet~\cite{pathak2022fourcastnet} have shown that data-driven methods can match or even surpass operational numerical weather prediction systems on forecasting benchmarks. However, these methods are black-box since they learn the dynamics purely from data without leveraging the well-understood physics of atmospheric transport. This can lead to physically implausible predictions, poor generalisation beyond the training regime ~\cite{wehenkel2023robust, naoya_2021}, and limited interpretability of the learned dynamics. Grey-box modelling~\cite{naoya_2021,verma2024climode,singh2026variational}, offers a principled middle ground. Even an incomplete but trusted physics model provides structural constraints, while a neural network learns the residual or unresolved processes from data.
Solver-based~\cite{naoya_2021,verma2024climode} grey-box models, however, face two limitations. First, training requires integrating the dynamics forward and backpropagating through every solver step \cite{chen2018neuralode}, incurring memory that scales linearly with the number of integration steps, longer training time and potential adjoint instability for stiff or chaotic dynamics. Second, the learned vector field is deterministic, while an incomplete physics model admits multiple plausible evolutions from near-identical atmospheric states; regressing a single field averages over these modes and blurs the learned dynamics \cite{guo2025variationalrectifiedflowmatching}.

In this work, we propose Climate Physics Dynamic Matching (ClimPhyDM), a variational simulation-free dynamics-informed framework for weather forecasting. Its dynamics
integrates a powerful atmospheric advection dynamics and a neural corrector adjusting systematic biases. Consecutive ERA5 states define target velocities via finite differences, against which the model's dynamics-composed vector field is regressed in a single forward pass, in the spirit of trajectory-level dynamics matching~\cite{zhang2024tjfm,singh2026variational}.
Stochasticity and multi-modality of the unresolved dynamics are captured by a latent variable inferred variationally from the history of states. Backpropagation through a solver is entirely avoided.

\begin{figure*}[t!]
\centering
\includegraphics[width=\textwidth]{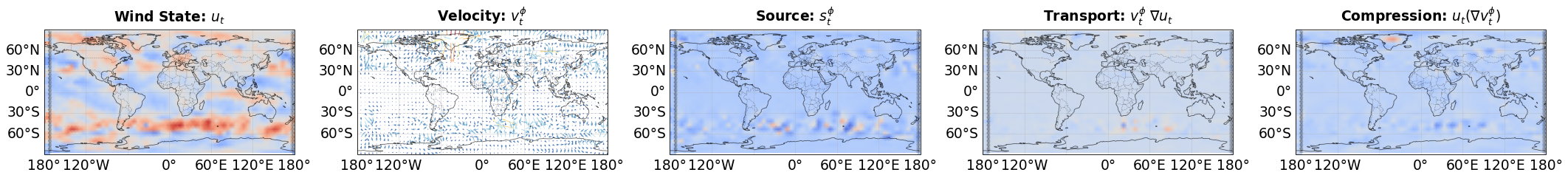}
\caption{\textbf{Vector field decomposition}. Term-wise visualization of ClimPhyDM's learned PDE components from Eq.~\eqref{eq:pde} at a single forecast step.}
\label{fig:grey_box_terms}
\end{figure*}

\section{Method}\label{sec:method}

\paragraph{\textbf{Climate Physics Dynamics.}}
Let $u_k \in \mathbb{R}^{C \times H \times W}$ denote the atmospheric state at discrete time $k$, where the $C$ channels correspond to meteorological variables and $H \times W$ is the latitude-longitude grid. We observe trajectories $\boldsymbol{u} = (u_0, u_1, \ldots, u_T)$ sampled from an unknown data distribution $\pi(\boldsymbol{u})$, and our goal is to learn a continuous-time model of the dynamics that, given the current state and a short history $\boldsymbol{u}_{k-h:k}$, forecasts future states. Following the physics of large-scale atmospheric transport \cite{verma2024climode}, we adopt an advection-type PDE as the physics prior. The evolution of the atmospheric field is governed by
\begin{small}
\begin{equation}
\frac{\partial u}{\partial t}
= \underbrace{-\,v^{\phi} \cdot \nabla u}_{\text{transport}}
\;\underbrace{-\;u \,(\nabla \cdot v^{\phi})}_{\text{compression}}
\;+\; \underbrace{s^{\phi}(u, \nabla u, z, \xi)}_{\text{source}},
\label{eq:pde}
\end{equation}
\end{small}
where the velocity field $v^ {\phi}(u, \nabla u, z, \xi)$ and source term $s^{\phi}$ are neural networks conditioned on the state, its spatial gradients, a stochastic latent $z$, and fixed spatio-temporal embeddings $\xi$; $s^{\phi}$ captures non-advective processes (diabatic heating, radiative forcing, subgrid turbulence). With the advection operator $f_p(u, v) = -\,v \cdot \nabla u - u\,(\nabla \cdot v)$, the grey-box vector field is
\begin{small}
\begin{equation}
V^{\phi}(u \mid z)
= f_p\big(u,\, v^{\phi}(u, \nabla u, z, \xi)\big)
+ s^{\phi}(u, \nabla u, z, \xi).
\label{eq:total-field}
\end{equation}
\end{small}
The physics acts as a structural constraint. The network predicts only the velocity passed through $f_p$, so transport and compression respect the continuity structure by construction, while the additive source retains flexibility for genuinely non-advective dynamics.

\paragraph{\textbf{Simulation-Free Variational Dynamics Matching.}}
Solver-based grey-box models such as ClimODE \cite{verma2024climode} train by integrating ~\eqref{eq:pde} and backpropagating through all $L$ solver steps, with $\mathcal{O}(L)$ memory and potential adjoint instability. We instead match dynamics directly \cite{zhang2024tjfm, singh2026variational}, for consecutive states $(u_k, u_{k+1})$ we form the interpolant $u_t = \frac{(t_{k+1}-t)\,u_k +  (t-t_k)\,u_{k+1}}{t_{k+1} - t_k}$, $t \in [t_k,t_{k+1}]$, whose derivative $\dot{u}_t = \frac{u_{k+1} - u_k}{t_{k+1} - t_k}$ is a supervised target for the vector field itself, so no solver enters training.

A deterministic field, however, averages over the multiple plausible velocities that an incomplete physics prior admits at a single state \cite{guo2025variationalrectifiedflowmatching}. We therefore model
 $p^{\phi}(\dot{u}_t \mid u_t) = \int p^{\phi}(\dot{u}_t \mid u_t, z)\, p(z)\, dz$ with prior $p(z)=\mathcal{N}(\mathbf{0},\mathbf{I})$, and introduce a variational posterior $q_{\psi}(z \mid \boldsymbol{u}_{k-h:k})$ conditioned on the recent segment, which disambiguates the active mode of the dynamics. With a Gaussian likelihood $p^{\phi}(\dot{u}_t \mid u_t, z) =
\mathcal{N}\big(\dot{u}_t; V^{\phi}(u_t \mid z), \sigma^2\mathbf{I}\big)$, the ELBO yields the objective
\begin{small}
    \begin{equation}
\mathcal{L}(\phi, \psi) =
\mathbb{E}_{\substack{\boldsymbol{u} \sim \pi,\; t \sim \mathcal{U}(t_k,t_{k+1})}}
\Big[
\mathbb{E}_{q_{\psi}(z \mid \boldsymbol{u}_{k-h:k})}
\big\| V^{\phi}(u_t \mid z) - \dot{u}_t \big\|^2
- \, \mathrm{KL}\big[ q_{\psi}(z \mid \boldsymbol{u}_{k-h:k}) \,\|\, p(z) \big]
\Big].
\label{eq:objective}
\end{equation}
\end{small}

where $t$ is drawn uniformly between the current state time $t_k$ and the next state time $t_{k+1}$, and $\boldsymbol{u}_{k-h:k}$ is the immediately preceding history of states.

\begin{figure*}[b!]
\centering
\includegraphics[width=\textwidth]{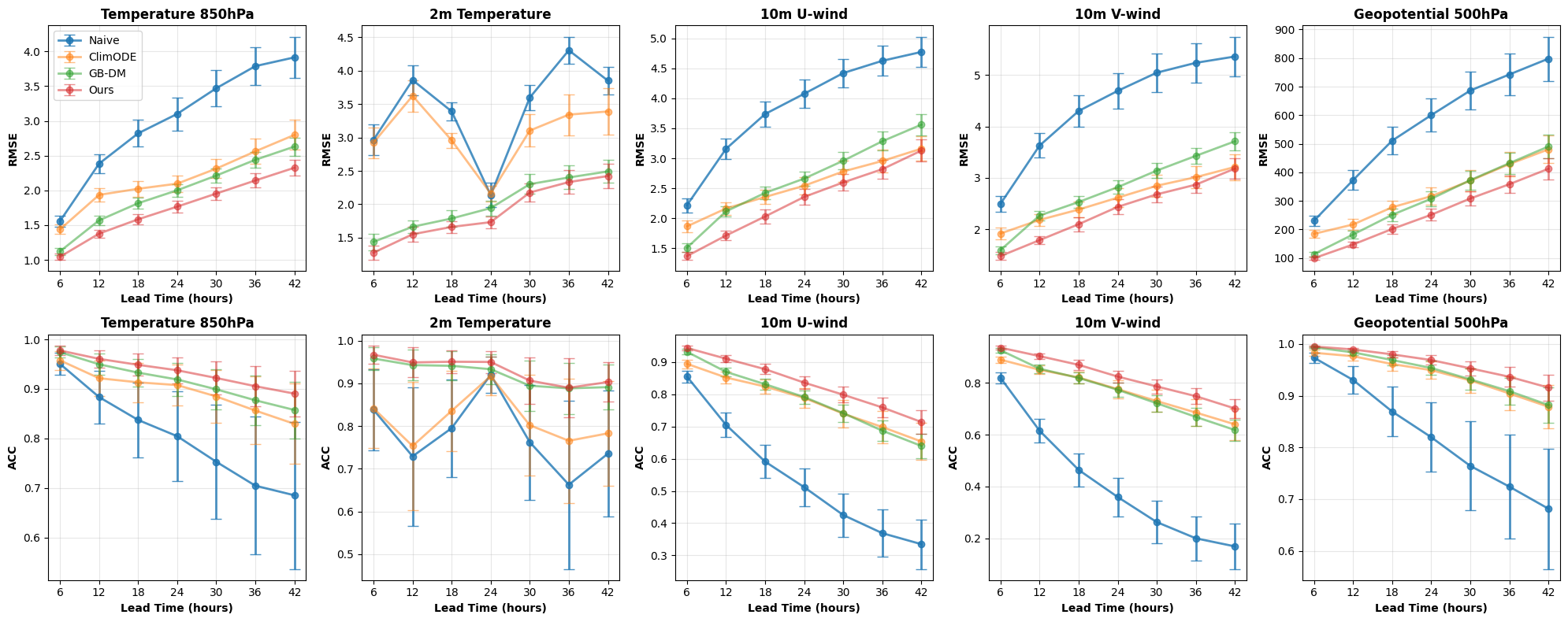}
\caption{\textbf{Hourly forecasting performance.} Latitude-weighted RMSE (top, lower is better) and ACC (bottom, higher is better) over 42 hours.}
\label{fig:clime_mse_global}
\end{figure*}

\section{Experiments}\label{sec:experiments}

\begin{figure*}[b!]
\centering
\includegraphics[width=\textwidth]{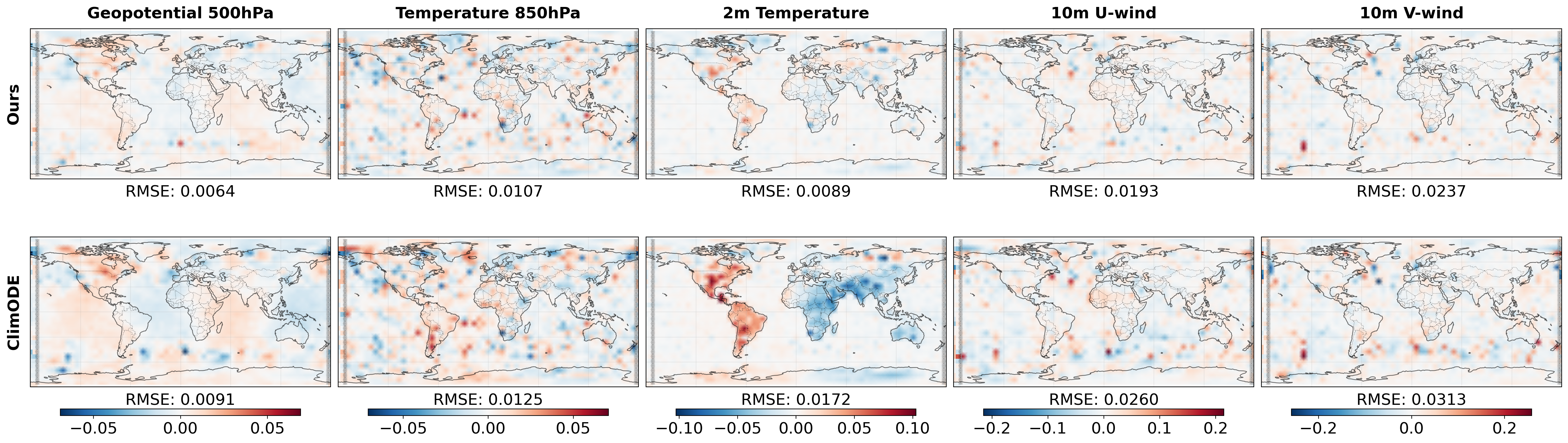}
\caption{\textbf{Residual maps for weather forecasting}. Visualization of absolute error maps between ground truth and predicted fields for five meteorological variables.}
\label{fig:residual_mse}
\end{figure*}

We train with the ERA5 reanalysis dataset as preprocessed by WeatherBench~\cite{rasp2020weatherbench} at $5.625^\circ$ spatial resolution and 6-hour temporal intervals, with five meteorological variables (geopotential at 500\,hPa (z500), temperature at 850\,hPa (t850), 2-metre temperature (t2m), and the 10-metre wind components (u10, v10)) min-max
normalised to $[0,1]$. Following Verma~\etal~\cite{verma2024climode}, we train on 2006-2015, validate on 2016, and test on 2017-2018.  We evaluate at two temporal resolutions: hourly, with forecasts up to 42 hours, and monthly, with forecasts up to 5 months.
We compare against ClimODE~\cite{verma2024climode}, the deterministic version of the proposed model GB-DM \cite{singh2026egu} and a persistence baseline (last observation carried forward). Comparing against GB-DM isolates the contribution of the variational latent, while ClimODE isolates the effect of simulation-free training.

\paragraph{\textbf{Hourly Resolution (42-Hour Horizon).}}
Figure~\ref{fig:clime_mse_global} reports latitude-weighted RMSE and ACC as functions of lead time. ClimPhyDM achieves lower RMSE and higher ACC than ClimODE and GB-DM, keeping temporal stability and resistance to error accumulation as the horizon increases. Figure~\ref{fig:residual_mse} corroborates this qualitatively. For a representative test sample, ClimPhyDM produces visibly lighter residual structures than both baselines on all variables, particularly over oceanic regions. Furthermore, our variational method enhances the deterministic baseline, highlighting its advantage over the deterministic GB-DM. Figure~\ref{fig:grey_box_terms} visualizes each learned PDE term, offering key interpretability of the model.

\begin{figure*}[b!]
\centering
\includegraphics[width=\textwidth]{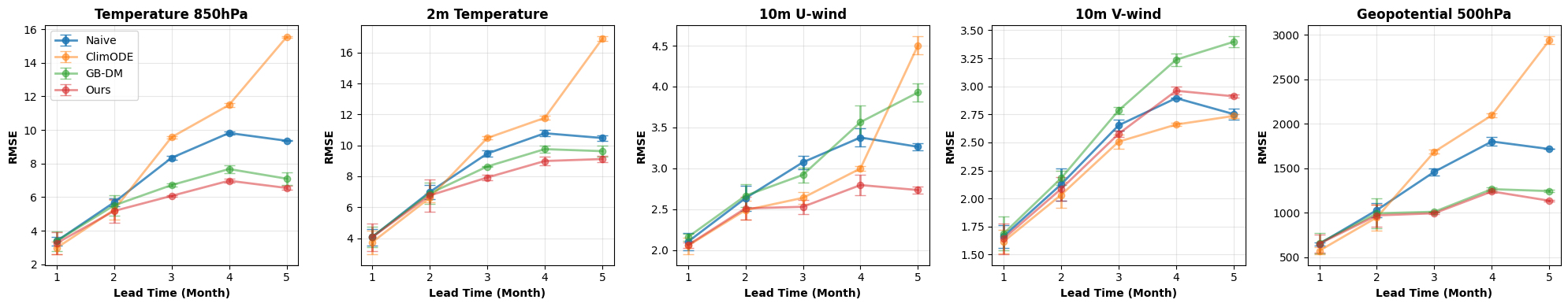}
\caption{\textbf{5-month forecast horizon.} Latitude-weighted RMSE (lower is better).}
\label{fig:clime_mse_month}
\end{figure*}

\paragraph{\textbf{Monthly Resolution (5-Month Horizon).}}
Figure~\ref{fig:clime_mse_month} extends the evaluation to a 5-month forecast horizon at monthly resolution. The gap to ClimODE widens with the lead time. While the two are close at short horizons, after the third month difference is substantial, indicating that our presents more consistent behaviour and accumulates error more slowly than solver-based training.

\section{Conclusion}
\label{sec:concl}
We introduced ClimPhyDM, a simulation-free grey-box framework that combines an advection-type physics prior with variational dynamics matching for weather forecasting. By regressing a physics-composed vector field directly onto finite-difference targets, training avoids the memory cost and
adjoint instability of solver-based primitive equations methods, while a variationally inferred latent captures multiple modalities of unresolved dynamics. On ERA5, ClimPhyDM outperforms ClimODE and its deterministic ablation at both hourly and monthly resolutions, with growing margins at longer horizons. Extensions to discontinuous regimes and higher resolutions are natural next steps.

\section*{Acknowledgements}
We acknowledge the financial support of the Swiss National Science Foundation (SNF). F. Lavda has been supported by SNF 200021\_207428, Learning Generative models for Molecules (LegoMol). G. S. Singh has been supported by the SNF project CRSII5\_209434, a Multidisciplinary and InteGRated Approach for geoThermal Exploration (MIGRATE). The computations were performed at the University of Geneva on ``Baobab'', ``Bamboo'' and ``Yggdrasil'' HPC clusters.

\bibliographystyle{splncs04}
\bibliography{bibliography}

\clearpage
\appendix
\nolinenumbers
\pagenumbering{arabic}
\setcounter{page}{1}
\renewcommand{\thefigure}{A\arabic{figure}}
\renewcommand{\thetable}{A\arabic{table}}
\renewcommand{\theequation}{A\arabic{equation}}
\begin{center}
   \begin{large}
    \textbf{APPENDIX}
\end{large}
\end{center}

\section{Background}\label{sec:background}
\subsection{Flow Matching for Dynamical Systems}\label{app:bg_fm}
Flow matching methods~\cite{lipman2023,albergo2023,liu2023flow} learn a time-dependent velocity field $v_t^\phi: [0,1] \times \R^d \to \R^d$ that transports samples from a source distribution $\pi_0$ to a target distribution $\pi_1$. Instead of solving the continuity equation through forward simulation, Conditional Flow Matching (CFM)~\cite{lipman2023} constructs a linear interpolation path $x_t = (1-t)x_0 + tx_1$ between paired samples $(x_0, x_1) \sim \pi(x_0, x_1)$, with target velocity $\dot{x}_t = x_1 - x_0$. The training loss is:
\begin{equation}
  \mathcal{L}_{\text{CFM}}(\phi) = \E_{x_0 \sim \pi_0,\, x_1 \sim \pi_1,\, t \sim \mathcal{U}(0,1)} \left\| v_t^\phi(x_t) - \dot{x}_t \right\|^2.
  \label{eq:cfm}
\end{equation}
This objective is simulation-free: it requires only a single forward pass through the network per training step, without numerical integration.

Trajectory Flow Matching (TFM)~\cite{zhang2024tjfm} extends CFM to dynamical systems where data arrives as time-series trajectories $\vect{x} = (x_0, x_1, \ldots, x_T)$. Rather than mapping noise to data, TFM constructs interpolation paths between consecutive trajectory states $x_k$ and $x_{k+1}$, learning the velocity field that governs the system's temporal evolution.
\subsection{Grey-Box Modelling}\label{app:bg_gb}
Grey-box modelling augments an incomplete physics model with a data-driven component. Given a physics model $f_p: \mathcal{X} \times \Theta \to \dot{\mathcal{X}}$ described by an ODE or PDE, the total dynamics are modelled as:
\begin{equation}
  \frac{\partial x_t}{\partial t} = f_\phi(x_t) \circ f_p(x_t, \theta),
  \label{eq:grey_box_general}
\end{equation}
where $f_\phi$ is a neural network and $\circ$ denotes function composition. The composition operator is deliberately more general than additive correction $f_\phi(x_t) + f_p(x_t, \theta)$: it allows the neural network to modulate the \emph{inputs} to the physics operator rather than simply correcting its output, thereby enforcing physical constraints on the learned dynamics.

Existing grey-box methods for dynamical systems~\cite{naoya_2021,yin2021augmenting,wehenkel2023robust,verma2024climode} typically train via Neural ODEs~\cite{chen2018neuralode}, integrating the combined dynamics forward with a numerical solver and backpropagating through the integration steps. While effective, this approach incurs $\mathcal{O}(L)$ memory and computational cost per training step, where $L$ is the number of solver steps.

\subsection{ELBO Derivation}
We report the full derivation of the ELBO used as the loss objective in \eqref{eq:objective} for training our conditional generative model.

\label{app:elbo-derivation}
\begin{align}
\log p_{t}^\phi(\dot{u}_t \mid u_k)
&= \log \int p_{t}^\phi(\dot{u}_t, z \mid u_k)\,  dz \\
&= \log \int q_{\psi}(z \mid \boldsymbol{u}_{k-h:k}) \frac{p_{t}^\phi(\dot{u}_t,  z \mid u_k)}{q_{\psi}(z \mid \boldsymbol{u}_{k-h:k})}\,  dz \\
&\geq \int q_{\psi}(z \mid \boldsymbol{u}_{k-h:k}) \log \frac{p_{t}^\phi(\dot{u}_t,  z \mid u_k)}{q_{\psi}(z \mid \boldsymbol{u}_{k-h:k})}\,  dz \quad \text{(Jensen's inequality)} \\
&= \int q_{\psi}(z \mid \boldsymbol{u}_{k-h:k}) \log \frac{p_{t}^\phi(\dot{u}_t \mid  z, u_k) p_t^\phi(  z \mid u_k)}{q_{\psi}(z \mid \boldsymbol{u}_{k-h:k})}\,  dz\\
&= \mathbb{E}_{q_{\psi}(z \mid \boldsymbol{u}_{k-h:k})}\left[\log p_{t}^\phi(\dot{u}_t \mid z, u_k) - \log \frac{q_{\psi}(z \mid \boldsymbol{u}_{k-h:k})}{p_t^\phi(  z \mid u_k)}\right] \\
&= \mathbb{E}_{q_{\psi}(z \mid \boldsymbol{u}_{k-h:k})}[\log p_{t}^\phi(\dot{u}_t \mid  z, u_k)] - \text{KL}\left[q_{\psi}(z \mid \boldsymbol{u}_{k-h:k}) \Vert p_t^\phi( z \mid u_k)\right]
\end{align}

Assuming a prior $p(z)$ independent of both data and time, i.e. :
\begin{equation}
   p_t^\phi( z \mid u_k) = p(z)
\end{equation}
    The objective becomes:
    \begin{align}
        \log p_{t}^\phi(\dot{u}_t \mid u_k) &\geq
        & \mathbb{E}_{q_{\psi}(z \mid \boldsymbol{u}_{k-h:k})}[\log p_{t}^\phi(\dot{u}_t \mid  z, u_k)] - \text{KL}\left[q_{\psi}(z \mid \boldsymbol{u}_{k-h:k}) \Vert p(z)\right]
    \end{align}

\section{Experimental details}

\subsection{Evaluation metrics}

Following~\cite{verma2024climode}, we assess benchmarks using latitude-weighted RMSE and Anomaly Correlation Coefficient (ACC) following the de-normalization of predictions.
$$
\text{RMSE} = \frac{1}{N}\sum_{t}\sqrt{\frac{1}{HW}\sum_{h}^{H}\sum_{w}^{W}\alpha(h)(y_{thw} - u_{thw})^{2}},$$

$$\text{ACC} = \frac{\sum_{t,h,w}\alpha(h)\tilde{y}_{thw}\tilde{u}_{thw}}{\sqrt{\sum_{t,h,w}\alpha(h)\tilde{y}_{thw}^{2}}\sqrt{\sum_{t,h,w}\alpha(h)u_{thw}^{2}}}
$$
where $\alpha(h) = \cos(h)/\frac{1}{H}\sum_{h}^{H}\cos(h')$ is the latitude weight and $\tilde{y} = y-C$ and $\tilde{u} = u - C$ are averaged against empirical mean $C = \frac{1}{N}\sum_{t}y_{thw}$.

\subsection{Dataset details}
We utilize the preprocessed ERA5 dataset provided by WeatherBench \cite{rasp2020weatherbench}, a widely used benchmark framework for evaluating data-driven weather forecasting models. The original ERA5 data at $0.25^\circ$ resolution is regridded by WeatherBench to coarser resolutions of $5.625^\circ$, $2.8125^\circ$, and $1.40625^\circ$; in this work, we adopt the $5.625^\circ$ dataset at 6-hour intervals. Our study focuses on $K=5$ key variables: 2-metre temperature (t2m), atmospheric temperature (t), geopotential (z), and the 10-metre wind vector components (u10, v10). All variables are normalized to the range $[0,1]$ using min--max scaling. Among these, $z$ and $t$ are standard verification variables in medium-range Numerical Weather Prediction (NWP) models, while t2m and (u10, v10) are directly relevant to surface-level conditions that impact human activities.

Following \cite{verma2024climode}, the dataset spans ten years of training data (2006--2015), one year for validation (2016), and two years for testing (2017--2018). The full set of ERA5 variables employed in our experiments is summarized in Table~\ref{tab:era5_vars}.

\subsection{Variational model and training}\label{app:variational}
Our grey-box model instantiates the velocity field $v^{\phi}$ and the source $s^{\phi}$ of Eq.~\eqref{eq:pde} as two residual convolutional networks -- a five/three/two-block ResNet for $v^{\phi}$ and a shallower two/one-block network for $s^{\phi}$ -- sharing the same input: the current state $u_t$, its spatial gradients $\nabla u_t$, cyclic time embeddings, and the static fields $\xi$. Convolutions use circular padding along longitude and reflection along latitude to respect the spherical grid, and a lightweight self-attention branch supplies global context. The variational posterior $q_{\psi}(z \mid \boldsymbol{u}_{k-h:k})$ is parametrised by a multi-head temporal attention encoder over a short history ($h=2$): the current state forms the query and the past states the keys and values, with attention applied only over time independently at each spatial location and learned position embeddings distinguishing the lags. The encoder emits the posterior mean and log-variance, from which $z$ is drawn by the reparameterisation trick, together with a deterministic context summary; both are concatenated to the inputs of $v^{\phi}$ and $s^{\phi}$.

\paragraph{\textbf{Training details.}} We optimise the ELBO of Eq.~\eqref{eq:objective} with AdamW at learning rate $5\times10^{-4}$, decayed by a cosine schedule to $10^{-5}$ and held constant thereafter, with gradient-norm clipping. The KL weight is annealed linearly from zero to $0.1$ to avoid posterior collapse early in training, and the matching target follows a stochastic interpolant whose noise level is annealed from $10^{-2}$ to $10^{-3}$, acting as a smoothing regulariser on the velocity field; a small penalty on the magnitude and spatial gradients of $v^{\phi}$ is applied early and decayed to zero. Numerical integration (an \texttt{rk4} solver) enters only at inference, never during training.

\begin{table}[h]
\centering
\caption{ECMWF data variables from ERA5 used in our dataset. \textit{Static} variables are time-independent, \textit{Single} represents surface-level variables, and \textit{Atmospheric} represents time-varying atmospheric properties at chosen altitudes.}
\label{tab:era5_vars}
\begin{tabular}{lllll}
\hline
Type & Variable name & Abbrev. & ECMWF ID & Levels \\
\hline
Static & Land-sea mask & lsm & 172 & \\
Single & 2 metre temperature & t2m & 167 & \\
Single & 10 metre U wind component & u10 & 165 & \\
Single & 10 metre V wind component & v10 & 166 & \\
Atmospheric & Geopotential & z & 129 & 500 \\
Atmospheric & Temperature & t & 130 & 850 \\
\hline
\end{tabular}
\end{table}

\subsection{Predictions visualization}
\begin{figure*}[b!]
\centering
\includegraphics[width=\textwidth]{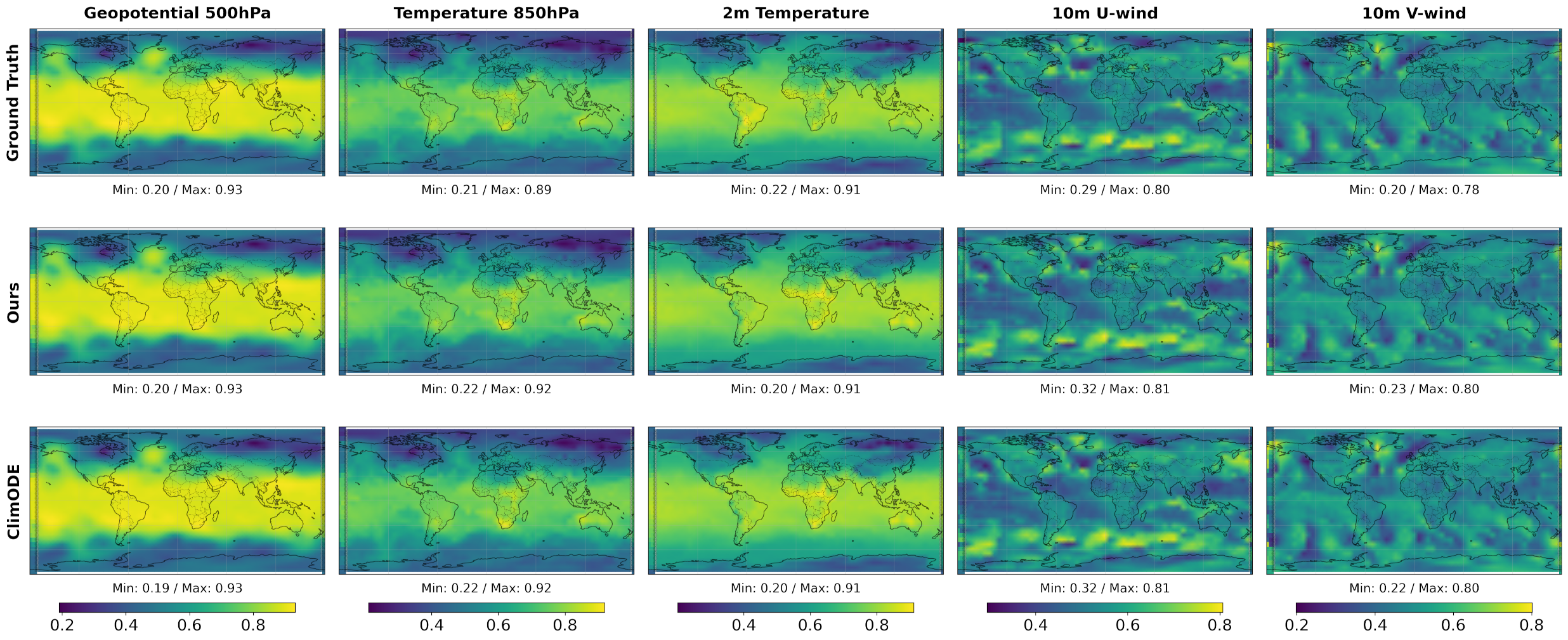}
\caption{Example of Climate forecasting using ClimPhyDM over $42$ hours.}
\label{fig:clime_predictions_hour}
\end{figure*}

In Figure~\ref{fig:clime_predictions_hour} we show the forecast of our model on the ERA5 dataset over 42 hours horizon.
In the main text we reported the error between true and predicted values to facilitate showing the difference between our model and ClimODE.

\section{Related Work}\label{app:related}

\paragraph{Data-driven weather forecasting.}
The past several years have seen rapid progress in data-driven weather prediction. Pangu-Weather~\cite{bi2023accurate} uses a 3D Earth-specific transformer trained on 39~years of ERA5 data. GraphCast~\cite{lam2023learning} employs graph neural networks on a multi-mesh representation. FourCastNet~\cite{pathak2022fourcastnet} leverages Fourier neural operators for global forecasting. GenCast~\cite{Price2023GenCastDE} applies conditional diffusion models for ensemble weather prediction. These methods achieve impressive forecasting skill but operate as black boxes, learning dynamics purely from data without leveraging known atmospheric physics. They may produce physically implausible predictions and offer limited interpretability of the learned dynamics.

\paragraph{Physics-informed neural networks.}
Physics-informed approaches incorporate physical knowledge into neural network training. PINNs~\cite{raissi2019physics} enforce PDE constraints via soft penalties in the loss function. Fourier Neural Operators~\cite{li2021fourier} learn mappings between function spaces with built-in resolution invariance. These methods incorporate physics as a regulariser rather than a structural component, and typically require access to the governing equations in full rather than working with incomplete physics models.

\paragraph{Grey-box dynamical systems.}
Grey-box methods that combine incomplete physics with neural networks have been explored across several domains. PhysVAE~\cite{naoya_2021} embeds physics ODEs into a variational autoencoder framework. \cite{yin2021augmenting} augment physical models with deep networks for complex dynamics. \cite{wehenkel2023robust} propose robust hybrid learning with expert augmentation. \cite{mehta2021neural} balance structure and flexibility in physical prediction. All these methods rely on Neural ODE solvers during training, inheriting the associated computational and stability limitations.

\paragraph{Flow matching and trajectory learning.}
Conditional Flow Matching~\cite{lipman2023,liu2023flow} provides simulation-free training for generative models. Trajectory Flow Matching~\cite{zhang2024tjfm} extends this to time-series data. Variational Rectified Flow Matching~\cite{guo2025variationalrectifiedflowmatching} introduces latent variables to handle multi-modal velocity fields. Our work adapts the trajectory flow matching paradigm to the grey-box setting, combining simulation-free training with compositional physics embedding.

\end{document}